\documentclass[aps,pra,floatfix,twocolumn,nofootinbib]{revtex4}

\usepackage{amsmath}
\usepackage{amssymb}
\usepackage{graphicx}
\usepackage{dcolumn}
\usepackage[sort&compress]{natbib}
\usepackage{bm}
\usepackage[dvips]{epsfig}

\usepackage{subfigure} 
\usepackage{float}

\usepackage{color}

\usepackage{graphicx} \usepackage{amsmath} \usepackage{amssymb}
\usepackage{subfigure}
\newcommand{\BEQ}{\begin{equation}}
\newcommand{\EEQ}{\end{equation}}
\newcommand{\BEA}{\begin{eqnarray}}
\newcommand{\EEA}{\end{eqnarray}}

\renewcommand{\a}{\alpha}

\newcommand{\g}{\gamma}

\newcommand{\bs}{{\bf s}}
\newcommand{\bu}{{\bf u}}
\newcommand{\bA}{{\bf A}}
\newcommand{\bh}{{\bf h}}

\newcommand{\at}{\tanh^{-1}}

\newcommand{\tq}{{\tilde q}}

\newcommand{\pin}{p_{\mathrm{in}}}
\newcommand{\pout}{p_{\mathrm{out}}}

\newcommand{\deltacrit}{\delta_{\mathrm{c}}}

\newcommand{\comment}[1]{}

\begin{document}

\title{Phase Transitions in Community Detection: A Solvable Toy Model}
\author{
Greg Ver Steeg$^1$, Cristopher Moore$^2$, 
Aram Galstyan$^1$\footnote{Corresponding author: galstyan@isi.edu},  and Armen E. Allahverdyan$^3$}
\affiliation{$^{1)}$USC Information Sciences Institute, Marina del Rey, CA, USA \\
$^{2)}$  Sante Fe Institute, Santa Fe, NM, USA\\
$^{3)}$Yerevan Physics Institute, Yerevan, Armenia
}

\begin{abstract} 
Recently, it was shown that there is a phase transition in the community detection problem. This transition was first computed using the cavity method, and has been proved rigorously in the case of $q=2$ groups.  However, analytic calculations using the cavity method are challenging since they require us to understand probability distributions of messages.  
We study analogous transitions in so-called ``zero-temperature inference'' model, where this distribution is supported only on the most-likely messages. Furthermore, whenever several messages are equally likely, we break the tie by choosing among them with equal probability. While the resulting analysis does not give the correct values of the thresholds, it does reproduce some of the qualitative features of the system. It predicts a first-order detectability transition whenever $q > 2$, while the finite-temperature cavity method shows that this is the case only when $q > 4$. It also has a regime analogous to the ``hard but detectable'' phase, where the community structure can be partially recovered, but only when the initial messages are sufficiently accurate. Finally, we study a semisupervised setting where we are given the correct labels for a fraction $\rho$ of the nodes. For $q > 2$, we find a regime where the accuracy jumps discontinuously at a critical value of $\rho$.
\end{abstract}

\maketitle

\section{Introduction}

A number of recent papers have studied fundamental limits on community detection in the stochastic block model (SBM), a simple generative model of networks with tunable modularity.  For networks that are dense enough, with an average degree that grows faster than $\log n$, the communities can be recovered exactly under some circumstances~\cite{Bickel-Chen-on-modularity}.  However, in the sparse case where the average degree is $O(1)$, there is a sharp transition below which the communities are undetectable~\cite{Reichardt2008,Allahverdyan2010a,Decelle2011PRL,Decelle2011PRE,Hu2012,Ronhovde2012,Nadakuditi2012}. The location of this transition was found using the cavity method~\cite{Decelle2011PRL,Decelle2011PRE}, or equivalently, by analyzing the behavior of the belief propagation (BP) algorithm. It was also hypothesized that BP is an optimal inference method for community detection in SBM, so that the corresponding detectability threshold is algorithm-independent~\cite{Decelle2011PRL,Decelle2011PRE}. 
This hypothesis was proved rigorously in the case of $q=2$ groups~\cite{mossel-neeman-sly,massoulie,mossel-neeman-sly-proof}; in the detectable regime, a polynomial-time algorithm exists that labels nodes correctly with probability bounded above $1/2$, while in the undetectable regime, graphs generated by the SBM are indistinguishable from Erd\H{o}s-R\'enyi random graphs, and no algorithm can label the nodes better than chance.

Below the transition, for $q \le 4$ belief propagation converges to a paramagnetic fixed point where every node is equally likely to belong to either community.  For $q > 4$, however, cavity method calculations~\cite{Decelle2011PRL,Decelle2011PRE} show that the situation is more complicated, including a ``hard but detectable'' regime where the communities can be recovered by belief propagation, but only if the algorithm is given a strong initial hint about the correct labels.  However, addressing this claim analytically is difficult, given that the cavity method requires us to keep track of an entire probability distribution of messages.

Here we study the community detection problem within the zero-temperature Bethe-Peierls approximation~\cite{Mezard1987,Mezard2001,Mezard2003}.  Equivalently, we study a message-passing algorithm where the distribution of messages is concentrated on the most likely label of each node.  Zero-temperature inference for community detection was also studied in~\cite{Reichardt2008}.  However, we augment this algorithm with a tiebreaking mechanism: Whenever a node has several equally-likely choices for its label, we break the symmetry randomly and uniformly among these choices, in effect applying an infinitesimal random external field.  This reduces the number of order parameters considerably, making it possible to study it analytically for any value of $q$.

We emphasize that the zero-temperature randomized message passing method should be thought of as a ``toy model'' for the real community detection problem. In particular, it overestimates the detectability thresholds, and the corresponding algorithm is far from optimal, as we discuss below.  Nevertheless, it reproduces some of the qualitative features of the real system and allows insight into the interesting but analytically difficult regime in which the graph contains many communities.

\paragraph*{Our Contributions.}  We  derive a fixed point equation for the proposed message-passing algorithm, and use it to analytically explore the community detection problem.  In this model, we find the detectability transition is continuous for $q=2$, and discontinuous for $q > 2$; in contrast, the finite-temperature cavity method~\cite{Decelle2011PRL,Decelle2011PRE} 
shows that the transition is continuous for $q \le 4$ and discontinuous for $q > 4$ (in the assortative case). 
Analogous to the ``hard but detectable'' regime~\cite{Decelle2011PRL,Decelle2011PRE}, we also find a regime in which there are two fixed points; a paramagnetic one where all labels are equally likely, and a second one which has high accuracy.  In this regime, the algorithm is able to recover the underlying community structure only if the initial messages are sufficiently close to the true labels; otherwise, it converges to the paramagnetic solution.

We also analyze the SBM reconstruction problem in ``semisupervised'' settings where one is given the true group labels for a fraction $\rho$ of nodes. For $q=2$, even a tiny amount of prior information suppresses the detectability transition in the zero-temperature model~\cite{Allahverdyan2010a}. Here we show that for $q > 2$, the behavior of the inference problem is much richer.  
Namely, while the prior information always improves the accuracy, there is a line of discontinuities where the accuracy jumps discontinously at a critical value of $\rho$, again in qualitative agreement with the cavity method for $q$ sufficiently large~\cite{zhang-moore-zdeborova}.

\section{Stochastic Block Models}

Consider a network of $N$ nodes, where each node belongs to one of $q$ groups or communities. Let $s_i \in \{1,\ldots,q\}$ be the community label of node $i$, and let $\bs=\{s_i\}_{i=1}^N$.  The probability of a link between two nodes $i, j$ in groups $k=s_i$ and $\ell=s_j$ is given by a $q \times q$ matrix $p_{k\ell}$.  We focus on the case where $p$ depends only on whether the nodes are in the same or different groups: that is, $p_{k\ell}=\pin \delta_{k\ell}+\pout (1-\delta_{k\ell})$.  We assume the network is \emph{assortative}, i.e., that $\pin > \pout$.  Finally, we assume it is sparse, i.e., that $\pin$ and $\pout$ are $O(1/N)$.

Let  $\bA$ be the adjacency matrix of a graph generated by the above block model. The generative model is fully described the following joint probability 
\BEQ
p(\bA,\bs)=\pi(\bs) p(\bA|\bs)=\pi(\bs)\prod_{i<j}p_{s_is_j}^{A_{ij}} (1-p_{s_is_j})^{1-A_{ij}}
\EEQ
where $\pi(\bs)$ encodes prior information about the community assignment one might have.

Given the observed network $\bA$, we are interested in reconstructing the unknown state $\bs$. Toward this goal, we define the posterior probability of $\bs$ given $\bA$, which, by the use of Bayes theorem, can be written as follows:  
\BEQ
p(\bs|\bA)=\frac{\pi(\bs) \,p(\bA|\bs)}{p(\bA)} \, . 
\EEQ
If the prior $p(\bA)$ is constant, this gives a Gibbs distribution $p(\bs|\bA) \propto \pi(\bs) \,p(\bA|\bs)$ at unit temperature.

There are several approaches for deciding the community assignments from the Gibbs distribution, and different approaches  are optimal for different loss functions~\cite{Iba1999}. For instance, 
the fraction of correctly inferred labels is maximized by computing marginal probabilities for each node, $p(s_i|\bA)$, and choosing the most-likely label for each one.  
Here we focus on a different approach known as \emph{maximum a posteriori} estimation that tries to find the state $\bs$ that jointly maximizes $p(\bs|\bA)$. This is the ground state of a generalized Potts Hamiltonian,
\BEQ
\mathcal H = -\sum_{i,j}A_{ij}\delta(s_i,s_j) + H_{\pi}(\bs)
\EEQ
where the second term represents prior knowledge about the community assignments.   

Since exact minimization is computationally intractable  for large graphs, one has to resort to approximate methods.  A popular family of such methods are message-passing algorithms such as belief propagation (BP).  
When the underlying graph is a tree, BP converges to the true marginals of the Gibbs distribution; although there are no convergence guarantees for general graphs with loops, the typical loop length in SBM scales as $\log N$, so we expect BP to be asymptotically correct in the thermodynamic limit~\cite{Decelle2011PRL,Decelle2011PRE}.  

If we want to find the ground state rather than the marginals, however, it makes sense to consider a zero-temperature version of belief propagation, 
where the messages are concentrated on the most-likely labels.  We describe this algorithm, and our simplification of it, in the next section.

\section{Zero-temperature Message Passing} 

In the zero-temperature form of belief propagation, also known as the max-product algorithm, the messages have a particular simple form. Namely, they are binary vectors $\bu=(u_1,u_2,\dots,u_q)$, where $u_k \in \{0,1\}$ for all $1 \le k \le q$ and at least one of the $u_k$ is positive. The message $\bu_{i\rightarrow j}$ (also referred to as cavity bias) from node $i$ to node $j$ describes the preferred state of node $i$ in the absence of node $j$.  To calculate this message, node $i$ sums the messages from all its neighboring nodes except $j$, obtaining the cavity field $\bh_{i \setminus j}=\sum_{k\neq j} A_{ki}\bu_{k\rightarrow i}$. It then constructs a new message $\bu_{i\rightarrow j}=\hat{\bu}(\bh_{i \setminus j})$, where the function $\hat{\bu}(\bh)$ picks the maximum component of its argument and sets it to $1$, while setting all the other components to  zero:  $\hat{u}_k(\bh)=\delta(h_k,\max_k h_k)$.  Thus $u_k(\bh_{i \setminus j})=1$ if $k$ is one of the most-likely groups for $i$ to belong to, given the most-likely group memberships of its neighbors other than $j$.

There are $2^q-1$ possible messages. Furthermore, the probability of  a particular message depends on the true label of the node it originates from, giving $q(2^q-1)$ probabilities.  However, due to symmetry, one can show that there are only $2q-1$ relevant order parameters~\cite{Reichardt2008}. Namely, what matters is (a) whether $u_k=1$ where $k$ is the correct label $s_i$, and (b) the number of other non-zero entries of $\bu$. Thus, the cavity field distributions can be parameterized as $\eta_{\ell,w}$, where $\ell=u_{s_i} \in \{0,1\}$, and $w=||\bu||^2-\ell \in \{0,\ldots,q-1\}$.

The fixed point of the message passing procedure can be found by solving the so-called cavity equation. For the SBM, this equation seems to have a closed-form solution only in simple cases, such as $q=2$~\cite{Reichardt2008,Allahverdyan2010a,Hu2012}. In general, one has to resort to numerical methods such as population dynamics.  Here one considers a pool of messages that are dynamically updated according to the rules specified above, while choosing the number of neighbors a node has in each group from the appropriate degree distribution.  In essence, this simulates the message-passing algorithm within the annealed approximation, where the network is redrawn at each iteration.

Here we modify the message-passing scheme by only allowing messages where exactly one of $\bu$'s components is $1$.  In our update, if the procedure above gives a message $\bu$ with more than one nonzero component, we break the tie by choosing one of those components with equal probability.  In that case, by symmetry and normalization, the only relevant order parameter is $m=\eta_{1,0}-\eta_{0,1}$, which we can think of as a magnetization.

\section{Analysis for $q=2$}

Below we use $\a=\pin N/q$ and $\g=\pout N/q$ to denote the average number of neighbors a node has in its own group and in each other group respectively.  The total connectivity, or average degree, is then $c=\a+(q-1)\g$.  We write $\delta = \a-\g$ as a measure of the strength of the community structure.

We start by analyzing zero-temperature message passing in the case $q=2$.  
There are three order parameters, which we denote $\eta_+=\eta_{10}$, $\eta_-=\eta_{01}$, and $\eta_0=\eta_{11}$, corresponding to correct, incorrect, and non-informative messages respectively.  The update rule is a majority vote, with $\eta_0$ corresponding to a tie.  Specifically, let $k_1$ be the number of correct messages $i$ receives from neighbors in its own group, plus the number of incorrect messages it receives from the other group; and let $k_2$ be the number of incorrect messages it receives from its own group, plus the number of correct messages from the other group.  If $k=k_1-k_2$, then $i$'s message is correct, incorrect, or uninformative if $k > 0$, $k < 0$, or $k=0$ respectively.

In networks generated by the stochastic block model, the number of neighbors a node has in its own group or in the other group are Poisson-distributed with mean $\a$ and $\g$ respectively (note that we can generalize this to other degree distributions).  As in the cavity method, we assume that the messages sent by these neighbors are independent.  Thus $k_1$ and $k_2$ are Poisson-distributed with mean $\lambda_1$ and $\lambda_2$ respectively, where
\BEQ
\lambda_1 = \a \eta_+ + \g \eta_- \, , \; 
\lambda_2 = \a \eta_- + \g \eta_+ \, . 
\EEQ
Their difference $k=k_1-k_2$ is then distributed according to the Skellam distribution:
\BEQ
P(k)= e^{-(\lambda_1 + \lambda_2)} \left( \frac{\lambda_1}{\lambda_2} \right)^{\!\frac{k}{2}} I_{|k|}(2\sqrt{\lambda_1 \lambda_2})
\label{eq:skellam}
\EEQ
where $I_k(z)$ is the modified Bessel function of the first kind. 
Without tiebreaking, the fixed-point equations of population dynamics are thus
\BEQ
\eta_+ = \sum_{k>0} P(k) \, , \; 
\eta_- = \sum_{k<0} P(k) \, .
\label{eq:without}
\EEQ
If we define 
\BEQ
m = \eta_+ - \eta_- \, , \; \tq = \eta_+ + \eta_- \equiv 1 - \eta_0 \, ,
\EEQ
then $m$ and $\tq$ are the magnetization and the Edwards-Anderson parameter, respectively. The fixed point equations can be rewritten in terms of these parameters,
\BEA
\label{eq:q} 1-\tq &=& e^{-c\tq} I_0(x) \\
m &=&  e^{-c\tq} \sum_{k=1}^{\infty} 
\left[ \left( \frac{\lambda_1}{\lambda_2} \right)^{\!\frac{k}{2}} - \left( \frac{\lambda_1}{\lambda_2} \right)^{-\frac{k}{2}} \right] I_{k}(x) 
\nonumber \\
&=& 2e^{-cq} \sum_{k=1}^\infty I_k(x) \sinh ky \, ,
\label{eq:m} 
\EEA
where we have defined
\BEA
\label{eq:x} x &\equiv& 2\sqrt{\lambda_1 \lambda_2} = \sqrt{c^2 \tq^2 - \delta^2 m^2} \\
\label{eq:y} y &\equiv& \at \frac{\lambda_1-\lambda_2}{\lambda_1+\lambda_2} = \at \frac{\delta m}{c\tq} \, . 
\EEA
These same equations were obtained via zero-temperature cavity methods in~\cite{Allahverdyan2010a}, 
but this derivation is considerably simpler. 
  
In the vicinity of the second-order phase transition we linearize~\eqref{eq:m} around $m=0$ to obtain
\BEQ
m = 2e^{-c\tq} \delta m \sum_{k=1}^\infty \frac{k I_k(c\tq)}{c\tq} \, . 
\label{che4}
\EEQ
Using the identity $2k I_k(z) / z = I_{k-1}(z)-I_{k+1}(z)$, the sum in~\eqref{che4} telescopes, giving an equation for the detection threshold,  
\BEQ
\deltacrit 
= \frac{e^{c\tq}}{I_0(c\tq) + I_1(c\tq)} \, ,
\label{thre1}
\EEQ
where $\tq$ is given by taking $m \to 0$ in~\eqref{eq:q},
\BEQ
\label{ha4}
1-\tq = e^{-c\tq} I_0(c\tq) \, . 
\EEQ

We now consider the case with tiebreaking, flipping a coin whenever $k=0$.  In that case~\eqref{eq:without} becomes
\BEQ
\eta_+ = \sum_{k>0} P(k) + \frac{1}{2} P(0) \, , 
\EEQ
and $\eta_+ + \eta_- = 1$.  After some manipulation, we obtain 
\BEQ
m =  2 e^{-c} \sum_{k=1}^\infty I_k(x) \sinh ky
\label{eq:mnew}
\EEQ
but where now $x$, $y$ are defined by 
\BEQ
x \equiv \sqrt{c^2 - \delta^2 m^2} \, , \; 
y \equiv \at \frac{\delta m}{c} \, .
\label{newx}
\EEQ
Reasoning as before, we obtain for the threshold
\BEQ
\deltacrit = \frac{e^c}{I_0(c) + I_1(c)} \, .
\label{thre_new}
\EEQ

\begin{figure}[htbp] 
   \centering
    \includegraphics[width=0.75\columnwidth]{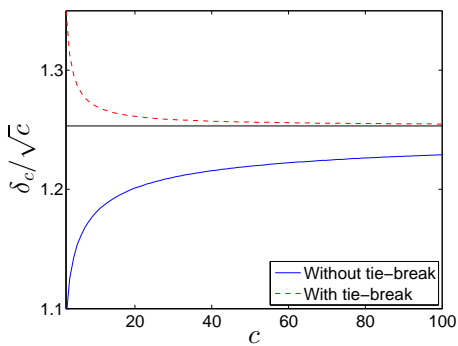} 
   \caption{Detection thresholds $\deltacrit$ for zero-temperature inference with and without tiebreaking, scaled by $\sqrt{c}$.  The true detectability threshold is at $\deltacrit/\sqrt{c}=1$; the zero-temperature thresholds converge to $\delta_c/\sqrt{c} = \sqrt{\pi/2}$ as $c \to \infty$.}
   \label{fig:thresholds_q=2}
\end{figure}

Comparing~\eqref{thre1} and~\eqref{thre_new} we see that tiebreaking is equivalent to setting the Edwards-Anderson parameter $\tq$ to $1$. In Fig.~\ref{fig:thresholds_q=2} we show both thresholds as a function of $c$.  The threshold with tiebreaking is higher, showing that it can be helpful to report ties rather than break them; this is reminiscent of distributed algorithms for approximate majority~\cite{AngluinAE2008majority}.  In fact, the tiebreaking algorithm fails to find communities even when $\gamma=0$, i.e., where \emph{all} links are within groups, if $c < 1.849$, since at this point $\delta_c=c$.
In contrast, without tiebreaking $\delta_c < c$ for all $c > 1$.

Neither version of zero-temperature inference performs as well as belief propagation. In particular, the detectability thresholds for both methods are noticeably higher than 
the algorithm-independent threshold predicted by~\cite{Decelle2011PRL,Decelle2011PRE} and established rigorously in~\cite{mossel-neeman-sly,massoulie,mossel-neeman-sly-proof}. Note, however, that those thresholds do scale correctly for large $c$: both~\eqref{thre1} and~\eqref{thre_new} approach $\deltacrit = \sqrt{(\pi/2) c}$ as $c \to \infty$, while the true threshold~\cite{Decelle2011PRL,Decelle2011PRE} is $\deltacrit = \sqrt{c}$.

\section{Analysis for Arbitrary $q$}

We now consider the case with tiebreaking for arbitrary $q$. By symmetry, we again have just two types of messages: correct ones with density $\eta_+$, and incorrect ones with density $\eta_- = 1-\eta_+$.  Incorrect messages are spread uniformly over the $q-1$ incorrect groups.

Let $k_0$ denote the number of messages a node $i$ receives carrying its own group label.  These are either correct messages from neighbors in its group, or a $1/(q-1)$ fraction of the incorrect messages from other groups.  The expected total number of neighbors $i$ has in other groups is $(q-1) \g$, so $k_0$ is Poisson with mean 
\BEQ
\lambda_1 
= \a \eta_+ + \gamma \eta_- 
= \gamma + \delta \eta_+ \, . 
\EEQ
For each of the other groups, which we label $\ell \in \{1,\ldots,q-1\}$, let $k_\ell$ be the number of messages $i$ receives with label $\ell$.  Then $k_\ell$ is Poisson with mean 
\BEQ
\lambda_2 
= \g \eta_+ + \frac{(\a + (q-2) \g) \eta_-}{q-1}
= \frac{(c-\gamma) - \delta \eta_+}{q-1} \, . 
\EEQ
The population dynamics then works as follows.  Let $k = \max \{k_0, k_1,\ldots, k_{q-1} \}$ and let $n = \sum_{\ell=1}^{q-1} \delta(k_\ell,k)$ be the number of \emph{incorrect} colors that achieve this maximum.  Then $i$ emits a correct message with probability $1/(n+1)$ if $k_0 = k$, and an incorrect message otherwise.

The joint probability that $n$ incorrect colors have $k_\ell=k$ and that the other $q-1-n$ have $k_\ell < k$ is
\BEQ
\label{barp}
\bar{p}(k,n) 
= 
{q-1 \choose n}
P_{\lambda_2}(k)^n \,Q_{\lambda_2}(k)^{q-1-n} \, ,
\EEQ
where $P_\lambda(k) = e^{-\lambda} \lambda^k / k!$ is the Poisson distribution with mean $\lambda$, and $Q_\lambda(k)= \sum_{j < k}  P_\lambda(j)$ is the regularized Gamma function.  The fixed point equation is then $\eta_+ = g(\eta_+)$ where
\BEQ
\label{geta}
g(\eta) = \sum_{k=0}^\infty P_{\lambda_1}(k) \sum_{n=0}^{q-1} \frac{\bar{p}(k,n)}{n+1} \, . 
\EEQ

\begin{figure}[t] 
   \centering
    \includegraphics[width=0.75\columnwidth]{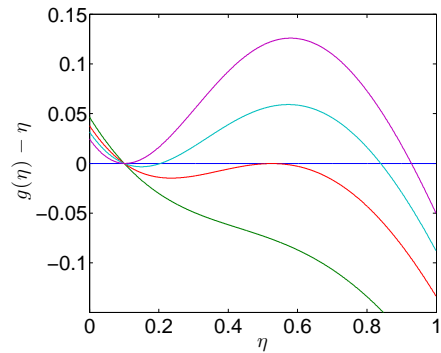} 
   \caption{Fixed point equation for $q=10$, $c=10$, and different values of $\delta$. From bottom to top, $\delta < \deltacrit^{(1)}$, $\delta = \deltacrit^{(1)}$, $\deltacrit^{(1)} < \delta < \deltacrit^{(2)}$, and $\delta = \deltacrit^{(2)}$. Observe that at $\deltacrit^{(1)}$, the second solution emerges discontinuously,  indicating a first order transition.}
   \label{fig:1}
\end{figure}

We illustrate the fixed point equation in Fig.~\ref{fig:1} for $q=10$ and $c=10$.  A close inspection reveals that there are several different phases separated by two phase transitions, which we denote $\deltacrit^{(1)}$ and $\deltacrit^{(2)}$. For $\delta < \deltacrit^{(1)}$, there is a single fixed point $\eta_+ = 1/q$ corresponding to the paramagnetic solution. At $\deltacrit^{(1)}$ a second solution $\eta_2 > 1/q$ emerges, giving an accurate labeling of the nodes.  This occurs when 
\BEQ
g(\eta_2)=\eta_2 \, , \; \left. \frac{dg}{d\eta} \right\vert_{\eta_2} = 1 \, .  
\EEQ
Similarly, $\deltacrit^{(2)}$ is defined by
\BEQ
\left. \frac{dg}{d\eta} \right\vert_{1/q} = 1 \, .  
\EEQ
For $q > 2$ this transition is first order; that is, $\eta_2$ is bounded above $1/q$.  In fact, the detectability transition (in the assortative case) is continuous for $q \le 4$ and first-order for $q > 4$~\cite{Decelle2011PRL,Decelle2011PRE}, but the zero-temperature model does give some intuition about why it becomes discontinuous at larger $q$.

The population dynamics can be described, in a suitable timescale, as $d\eta_+/dt = g(\eta_+) - \eta_+$.  
Therefore, for $\deltacrit^{(1)} < \delta < \deltacrit^{(2)}$, both $1/q$ and $\eta_2$ are locally stable, with an unstable fixed point between them.  
At $\deltacrit^{(2)}$, the paramagnetic solution $1/q$ becomes unstable.  
\begin{figure}[t] 
   \centering
    \includegraphics[width=0.75\columnwidth]{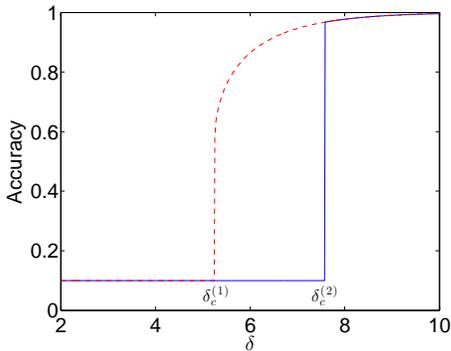} 
    \caption{The accuracy $\eta_+$ as a function of $\delta$ for $q=10$ and $c=20$.  The thresholds $\deltacrit^{(1)}$ and $\deltacrit^{(2)}$ correspond to the appearance of the second solution $\eta_2 > 1/q$ and the instability of the paramagnetic solution respectively.  The dashed line corresponds to initial messages accurate enough to converge to $\eta_2$, and the solid line corresponds to random initial messages. Compare Fig.~2(c) in~\cite{Decelle2011PRL}.}
        \label{fig:2}
\end{figure}

These results fit qualitatively with the results from the cavity method in~\cite{Decelle2011PRL,Decelle2011PRE}, albeit with overestimated values of $\deltacrit^{(1)}$ and $\deltacrit^{(2)}$.  If $\delta < \deltacrit^{(1)}$, the communities are undetectable, since the algorithm converges to the paramagnetic fixed point.  If $\delta > \deltacrit^{(2)}$, the communities are easy to detect, since a small perturbation away from the paramagnetic fixed point will lead to $\eta_2$.  Finally, $\deltacrit^{(1)} < \delta < \deltacrit^{(2)}$ is the ``hard but detectable'' regime: we can converge to $\eta_2$ and label the nodes accurately, but only if the initial messages are accurate enough.

In Fig.~\ref{fig:2} we plot the accuracy $\eta_+$ as a function of $\delta$.  The two curves correspond to different ways to initialize the messages; randomly (solid) and accurately enough to converge to $\eta_2$ (dashed). The gap between the two transitions corresponds to the regime where the non-paramagnetic solution exists but is hard to find. 

\begin{figure}[t] 
   \centering
    \includegraphics[width=0.75\columnwidth]{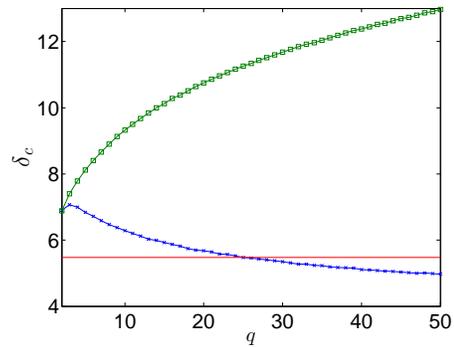} 
    \caption{Zero-temperature thresholds $\deltacrit^{(1)}$ (blue) and $\deltacrit^{(2)}$ (green) as a function of the number of groups $q$ for $c=30$.  The red line shows the true hard/easy threshold $\sqrt{c}$.}
        \label{fig:3}
\end{figure}
In Fig.~\ref{fig:3} we plot the thresholds $\deltacrit^{(1)}$ and $\deltacrit^{(2)}$ as a function of $q$, while keeping $c$ fixed.  While belief propagation succeeds whenever $\delta$ is above the true easy/hard transition $\sqrt{c}$ for any $q$, it appears that $\deltacrit^{(2)}$ increases with $q$.  
This suggests that, when starting from random messages, zero-temperature inference with tiebreaking performs poorly when the number of communities is large.

\section{Semisupervised Inference} 

So far, we have assumed that the only information available to us is the graph structure.  We now focus on ``semisupervised'' inference, where we are also given some prior information about the true group assignment.  

One can distinguish two possible scenarios.  In the first, we have \emph{noisy} information about every node, biasing us toward its correct label.  We can represent this by giving the correct label some weight $\beta > 1$ in the tiebreaking rule; then the probability $1/(n+1)$ of a correct message in~\eqref{geta} becomes $\beta/(n+1)$.

In the second scenario, we have information that is \emph{perfectly accurate}, but \emph{limited}: namely, we know the true labels of a fraction $\rho$ of the nodes~\cite{Allahverdyan2010a}.  Here we define $\eta_+$ as the accuracy we achieve on the unknown nodes.  In that case, we can modify our previous analysis by assuming that a fraction $\rho$ of the incoming messages are from known nodes, and are automatically correct.  Thus we replace $\lambda_1$ and $\lambda_2$ in~\eqref{barp} and~\eqref{geta} with 
\BEQ
\lambda_1(\rho) = \rho \a + (1-\rho) \lambda_1 \, , \;
\lambda_2(\rho) = \rho \g + (1-\rho) \lambda_2 \, . 
\EEQ
We found that these two scenarios produce qualitatively similar results, and we focus on the latter one. 

\begin{figure*}[!] 
   \centering
   \subfigure{
    \includegraphics[width=0.7\columnwidth]{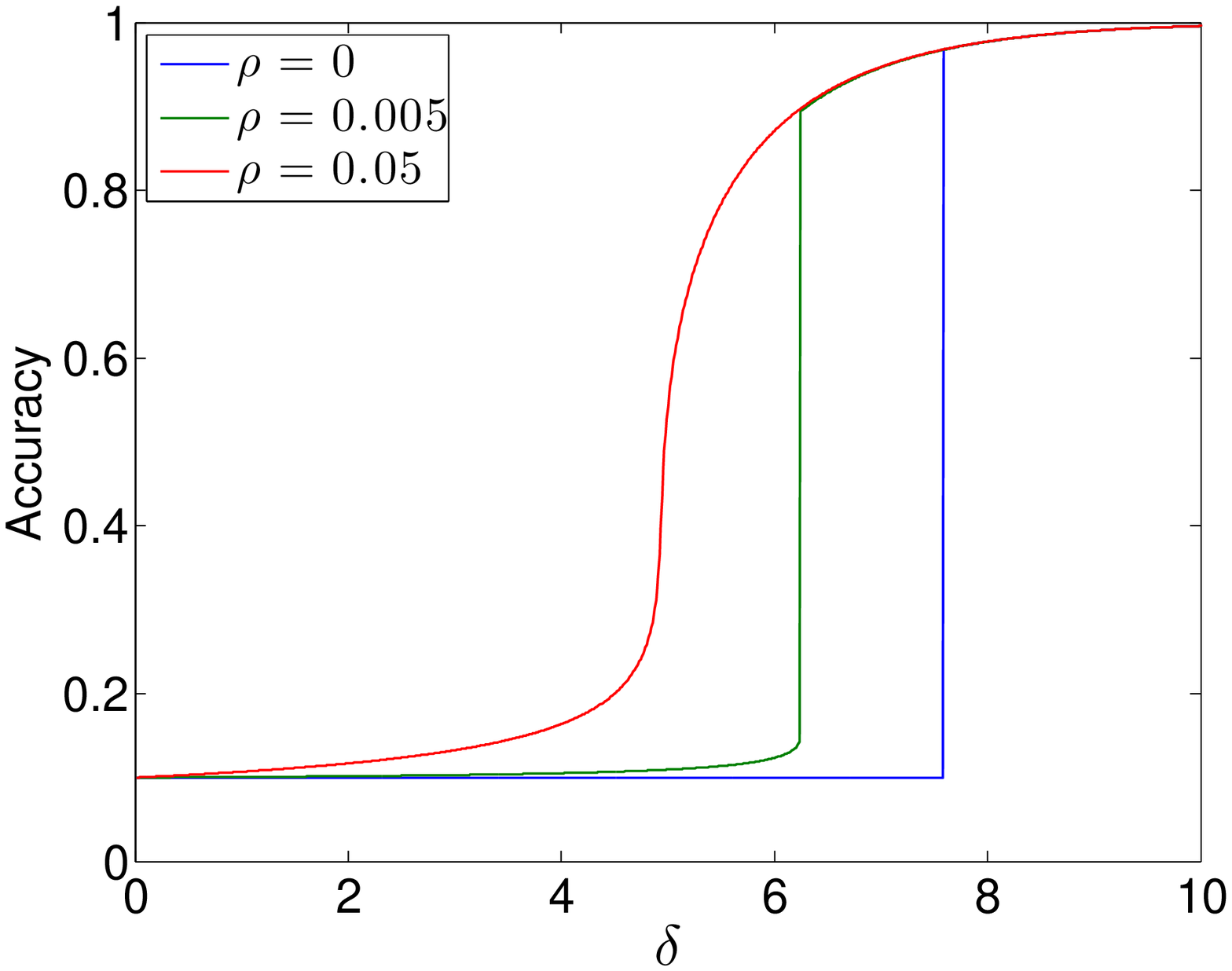} 
    \label{accuracy-ss}
    }
    \subfigure{
    \includegraphics[width=0.7\columnwidth]{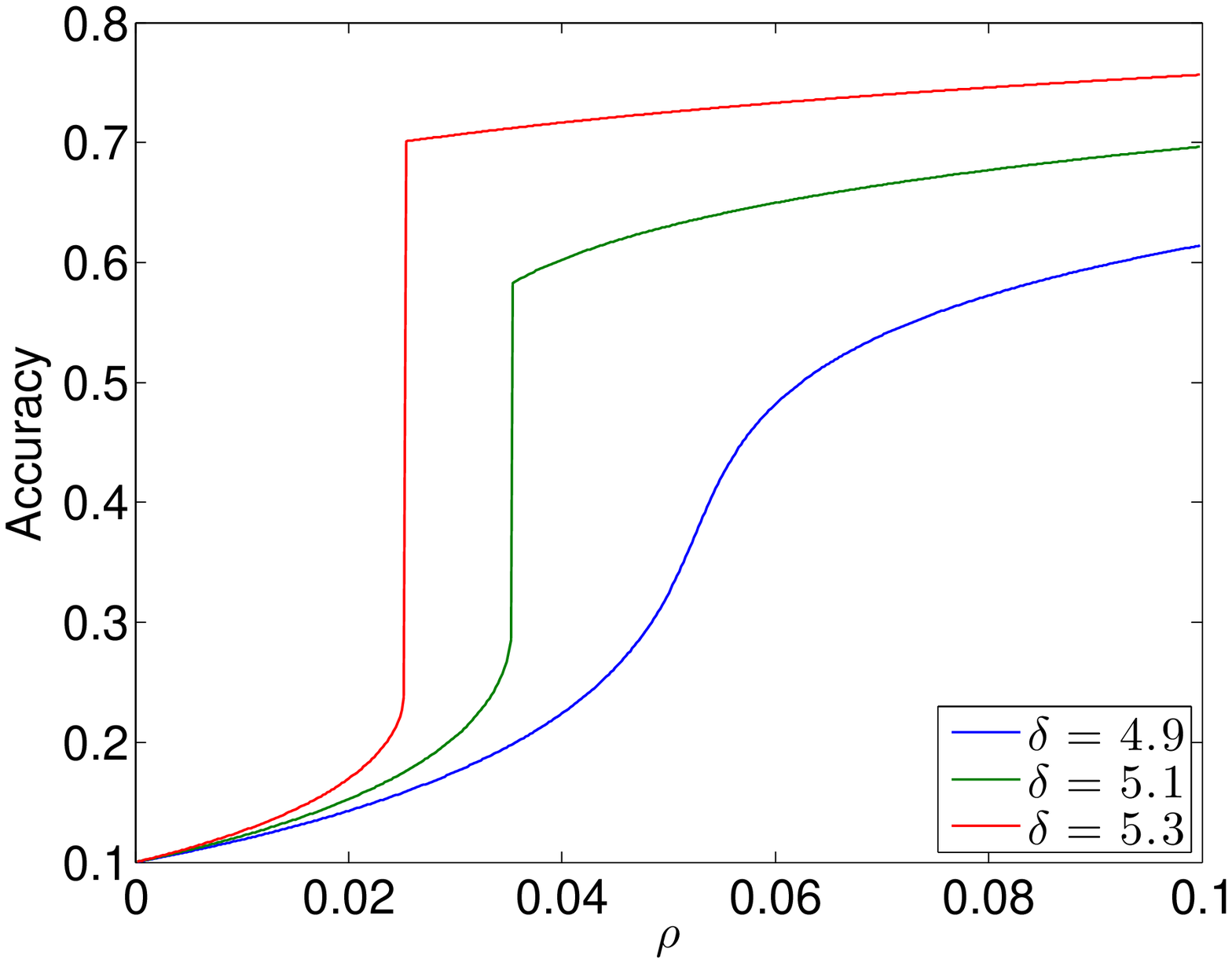} 
    \label{accuracy_vs_rho}
    }
   \caption{(a) Accuracy $\eta_+$ vs.\ $\delta$ for $q=10$ and $c=20$, and for varying amounts of prior information, $\rho=0.0,0.005,0.05$.  At $\rho=0$, we do no better than chance until $\delta = \deltacrit^{(2)}$, as in Fig.~\ref{fig:2}.  However, even a small $\rho$ moves the boundary betwen the easy and hard regimes downward, letting us jump to an accurate fixed point.  (b) Accuracy vs.\ $\rho$, for the same parameters as in (a), and for three different values of $\delta$.  There is a range of $\delta$ where the accuracy jumps discontinuously at a critical value of $\rho$.  At a critical value of $\delta$, this discontinuity disappears.}
\end{figure*}

Fig.~\ref{accuracy-ss} shows the accuracy $\eta_+$ as a function of $\delta$ for different amounts of prior information, $\rho=0, 0.005, 0.05$.  We see that even a small value of $\rho$ lets us jump to an accurate fixed point analogous to $\eta_2$ at some $\delta < \deltacrit^{(2)}$, letting us label the nodes even when we are some distance inside the ``hard but detectable'' regime.  Note that this observation is in stark contrast with the behavior reported in Ref.~\cite{Allahverdyan2010a}, where the detection threshold disappeared for any finite positive $\rho$. 

As shown in Fig.~\ref{accuracy_vs_rho}, there is a range of $\delta$ where the accuracy jumps discontinuously at a critical value of $\rho$.  These discontinuities disappear at a particular value of $\delta$, correponding to a tricritical point.  Below this $\delta$ the accuracy increases steeply, but continuously, as a function of $\rho$.  This qualitatively reproduces the picture from cavity method calculations     for large $q$~\cite{zhang-moore-zdeborova}.

\section{Discussion}

We analyzed community detection in the stochastic block model, based on a zero-temperature message-passing algorithm.  By breaking ties randomly, we reduced the number of order parameters to one, giving us an analytically tractable model for any number of groups.

The randomized message passing algorithm considered here is not optimal for the community detection problem. Therefore, any detection thresholds reported here can only be viewed as bounds on the true (algorithm-independent) detection thresholds. Nevertheless, it lets us analytically reproduce some qualitative aspects of the true transition.  For $q > 2$ it predicts a first-order detectability transition, and a ``hard but detectable'' regime.  We note that the finite-temperature cavity method shows that this regime appears when $q > 4$~\cite{Decelle2011PRL,Decelle2011PRE}.  

We also analyzed a ``semisupervised'' setting where one is given the true labels of a fraction $\rho$ of nodes. In contrast to $q=2$~\cite{Allahverdyan2010a}, for $q > 2$ even a small value of $\rho$ significantly moves the boundary between the hard and easy regimes, and there is a range of $\delta$ where the accuracy jumps discontinuously as a function of $\rho$.  This is again in qualitative agreement with the cavity method~\cite{zhang-moore-zdeborova}.  

We limited our analysis to the case where the connectivity between nodes depends only on whether they are in the same group or not, and where the groups are of equal size.  Our approach can be generalized to more general cases, although the analysis will be more complicated.

\acknowledgments
A.G. and G.V.S. thank the Santa Fe Institute for their hospitality.  A.G. and G.V.S. were supported in part by the US AFOSR MURI grant FA9550-10-1-0569, and US DTRA grant HDTRA1-10-1-0086.  C.M. is supported by the AFOSR and DARPA under grant \#FA9550-12-1-0432.  We thank Lenka Zdeborov\'a, Pan Zhang, Elchanan Mossel, and Allan Sly for helpful conversations.


\end{document}